\documentclass[aps,prl,reprint,twocolumn,groupedaddress]{revtex4-1}

\usepackage{graphicx}
\usepackage{textcomp}
\usepackage{upgreek}
\usepackage{amssymb}
\usepackage[version=3]{mhchem}
\usepackage[shortcuts]{extdash}

\begin{document}

\title{On the breakdown of the simple Arrhenius law in the normal liquid state}

\author{Erik Thoms}
\email[]{erik.thoms@smcebi.edu.pl}

\author{Andrzej Grzybowski}

\author{Sebastian Pawlus}

\author{Marian Paluch}
\email[]{marian.paluch@us.edu.pl}
\affiliation{Institute of Physics, University of Silesia, Uniwersytecka 4, 40-007 Katowice, Poland}
\affiliation{Silesian Center for Education and Interdisciplinary Research, 75 Pulku Piechoty 1A, 41-500 Chorzow, Poland.}

\date{\today}

\begin{abstract}
	It is common practice to discuss the temperature effect on molecular dynamics of glass formers above the melting temperature in terms of the Arrhenius law.
	Using dielectric spectroscopy measurements of dc-conductivity and structural relaxation time on the example of the typical glass former propylene carbonate, we provide experimental evidence that this practice is not justified.
	Our conclusions are supported by employing thermodynamic density scaling and the occurrence of inflection points in isothermal dynamic data measured at elevated pressure.
	Additionally, we propose a more suitable approach to describe the dynamics both above and below the inflection point based upon a modified MYEGA model.
\end{abstract}

\maketitle

	Glasses, \textit{i.e.} solids in a non-crystalline state, and supercooled liquids are substantial materials for a multitude of technical applications.
	From a physical point of view, however, the glass transition process is still not fully explored and many aspects and characteristics wait to be explained satisfactorily. 
	In particular, the roles of the free volume on the one hand and thermal energy on the other hand are not yet determined.
	
	A common way to transfer a material into the glassy state is supercooling. 
	By decreasing the temperature under the melting point $T_m$ at a rate high enough to avoid crystallisation, the viscosity $\eta$ of the liquid can be increased, until a solid without long range order forms below the glass temperature $T_g$.
	One characteristic property of glass formers in the supercooled state is the non-Arrhenius dependency of $\eta(T)$ upon cooling. 
	A simple energy barrier activated behavior is given by the Arrhenius equation
	\begin{equation}
		\eta (T) = \eta_{\infty} \times \textnormal{exp}\left(\frac{E_{act}}{RT}\right),
		\label{equ-arrh}
	\end{equation}
	where $T$ is the absolute temperature, $\eta_{\infty}$ the viscosity for very high $T$, and $R$ the universal gas constant. 
	In contrast, a commonly employed model to describe the super-Arrhenius temperature dependence observed in glass formers is the empirical Vogel-Fulcher-Tammann-Hesse equation (VFTH equation) \cite{Vogel1921,Fulcher1925,Tammann1926}:
	\begin{equation}
		\eta (T) = \eta_{\infty} \times \textnormal{exp}\left(\frac{D_{VF} T_{VF}}{T-T_{VF}}\right)
		\label{equ-vf}
	\end{equation}
	Here, $T_{VF} < T_{g}$ is the Vogel-Fulcher temperature, indicating a divergence point, while the strength parameter $D_{VF}$ quantifies the discrepancy from ideal Arrhenius behavior.
	In the case of other properties, \textit{e.g.}, structural relaxation time $\tau_{\alpha}$ or inverse dc-conductivity $\sigma_{dc}^{-1}$, a similar temperature dependence can be observed and be described by analogous models. 
	
	Besides supercooling, the generation of a vitreous state is also possible by limiting the free volume of the molecules via an increase of the pressure $p$.
	Close to the glass transition, a pressure equivalent of the VFTH-law can be used to describe the non-activation pressure $p$-dependence of $\eta$ and other quantities \cite{Paluch1999}:
	\begin{equation}
		\eta (p) = \eta_0 \times \textnormal{exp} \left(\frac{C_{F} p}{p_{\infty} - p} \right)
		\label{equ-pvf}
	\end{equation}
	$\eta_0 = \eta (p \rightarrow 0)$ is the viscosity at very low pressures, $p_{\infty}$ is the divergence pressure and $C_{F}$ is a pressure equivalent to $D_{VF}$.
	At conditions far from the glass transition (\textit{i.e.} $T \gg T_{g}(p)$), however, deviation from this behavior is observed and sometimes described by the pressure counterpart of the Arrhenius law,
	\begin{equation}
		\eta (p) = \eta_{0} \times \textnormal{exp}\left( \frac{p V_{act}}{RT} \right),
		\label{equ-parrh}
	\end{equation}
	with the constant activation volume $V_{act}$ \cite{Paluch2015}.
	While an isobaric change of temperature effects both the available energy and the molecular packing, isothermal modification of pressure only acts on the latter. 
	As a consequence, it is possible to separate the influence of those two factors by a systematic analysis of the pressure- and the temperature-behavior of a glass former \cite{Floudas2011}, which makes high pressure studies a key element for understanding the glass transition.
	
	Despite its prevalence, the VFTH-model met with criticism, in particular, due to the lack of experimental evidence for a divergence of $\tau_{\alpha}$ at finite temperatures, and the resulting break down of the model at low temperatures \cite{Laughlin1972,Hecksher2008}. 
	Based on the Adam-Gibbs model \cite{Adam1965} and the Phillips-Thorpe constraint theory \cite{Phillips1979,Thorpe1983}, Mauro \textit{et al.} proposed a three parameter entropy based model avoiding such divergence \cite{Mauro2009}, that has been tested successfully by other groups \cite{Lunkenheimer2010,Garcia2011}.
	According to this MYEGA-model, the temperature dependence of dynamic properties can be described by 
	\begin{equation}
		\eta (T) = \eta_{\infty} \times \textnormal{exp}\left[\frac{K}{T}\textnormal{exp}\left(\frac{C}{T}\right)\right],
		\label{equ-myega}
	\end{equation}
	with $K$ being related to the activation energy and $C$ to the energy difference between the intact and broken states of network constraints.

	One of the most interesting findings in recent years in the study of glass forming materials is thermodynamic density scaling.
	For several supercooled liquids, it has been reported that a scaling function $J(\Gamma)$ can describe the thermodynamic dependencies of dynamic properties, \textit{e.g.}, the isothermal and isobaric structural relaxation times $\tau_{\alpha}$, close to the glass transition \cite{Toelle2001,Dreyfus2003,Alba-Simionesco2004,Casalini2004,Roland2005,Floudas2011-2}.
	Consequently, the measured data for multiple $p$ and $T$ values collapse to one master curve when plotted against $\Gamma$.
	The scaling variable $\Gamma$ can be expressed by temperature $T$ and specific volume $V$ via
	\begin{equation}
		\Gamma = T^{-1} V^{-\gamma},
		\label{equ-Gamma}
	\end{equation}
	where the scaling exponent $\gamma$ is a material constant independent of thermodynamic conditions, and
	\begin{equation}
		\Gamma = \textnormal{const. for } \tau_{\alpha} = \textnormal{const.}
		\label{equ-scaling}
	\end{equation}
	(respectively $\sigma_{dc} = \textnormal{const.}$) is the scaling criterion \cite{Paluch2007}.
	These findings sparked high interest, as they potentially could bring light to relations between thermodynamics and relaxation dynamics.\par 
	
	Conforming to the scaling criterion, Masiewicz \textit{et al.} generalised the MYEGA model to extend its applicability to $T-p$ dependent data transformed via $V=V(T,p)$ to the $T-V$ domain \cite{Masiewicz2012}:
	\begin{equation}
		\tau_{\alpha} = \tau_{0} \times \textnormal{exp} \left[ \frac{D}{T V^{\gamma}}\textnormal{exp} \left( \frac{A}{T V^{\gamma}} \right) \right].
		\label{equ-myegatv}
	\end{equation}
	The relaxation time $\tau_{0}$ for high $T V^{\gamma}$, \textit{i.e.} $\Gamma \rightarrow 0$, as well as $D$ and $A$ are fitting parameters.

	In this paper, we employ the scaling criterion to generate pressure dependent $\tau_{\alpha}$ data from ambient pressure dielectric spectroscopy measurements on the example of the canonical glass former propylene carbonate (PC).
	The generated data is verified \textit{via} comparison with measured dc-conductivity data, and is fitted according to a modified MYEGA model.
	We find an inflection point in the pressure dependent curves, separating distinctly different $p$-dependency at low and high pressure ranges.
	Using various calculated $\tau_{\alpha}$ isotherms, we analysed the temperature and relaxation time dependences of the inflection point. 

	The propylene carbonate was purchased from Sigma-Aldrich with a purity $>99\%$.
	Details on the experimental setup can be found in a earlier work \cite{Paluch2003}.
	
 \begin{figure}[tb]
  \centering
  \includegraphics[width=8.5cm]{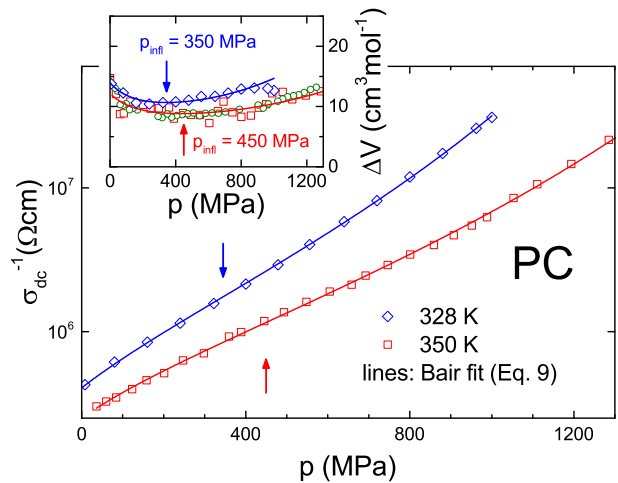}
	\caption{Measured pressure dependent inverse dc-conductivity of PC for a constant temperature of 328\,K (diamonds) respectively 350\,K (squares). In the inset, the corresponding activation volumes are shown, together with the activation volume derived from the structural relaxation times (circles) given in Fig. \ref{fig-ds}(b). The lines are fits according to Eq. \ref{equ-hybrid}.}
  \label{fig-meas}
 \end{figure}

	In Fig. \ref{fig-meas}, the inverse dc-conductivity $\sigma_{dc}^{-1} (p)$ is shown for $T=328$\,K and 350\,K in a pressure range up to $p \approx 1380$\,MPa. 
	For the 350\,K isotherm, the conductivity decreases from a value of $\sigma_{dc} \approx 4 \times 10^{-6} \textnormal{ } \textnormal{S}/\textnormal{cm}$ at ambient pressure to approximately $3 \times 10^{-8} \textnormal{ } \textnormal{S}/\textnormal{cm}$ over this range.
	Generally, the decrease of molecular mobility with rising pressure induces an increase in $\sigma_{dc}^{-1}$.
	It can be seen from the pictured data that the pressure dependence $\sigma_{dc}^{-1} (p)$ changes from a slower than exponential growth, \textit{i.e.} negative curvature respectively concave shape versus the $p$-axis in Fig. \ref{fig-meas} at low pressures to faster than exponential growth at elevated pressures (positive curvature and convex shape) \cite{Bair2015}.
	Consequently, the low pressure range is not well described by a pressure dependent Arrhenius-like law (Eq. \ref{equ-parrh}).
	The 328\,K isotherm shows equal behavior, albeit at generally lower conductivity.
	Similar results have recently been observed in the conductivity of protic ionic liquids \cite{Thoms2017}, as well as in the structural relaxation time $\tau_{\alpha}$ or the viscosity $\eta$ of molecular or ionic liquids \cite{Cook1994,Casalini2008,Gacino2015}, even though the concave progression at low pressures is not necessarily incorporated in the models used in these reports.
	
	The presence of the inflection point becomes more obvious when the $\sigma_{dc}(p)$ data is analyzed in terms of the activation volume parameter $\Delta V_{act,\sigma}(p) = RT \times \left(\partial \textnormal{ln}(\sigma_{dc}^{-1})/\partial p \right)$ as presented in the inset of Fig. \ref{fig-meas}. 
	In this representation, slower and faster than exponential pressure dependences correspond to a decrease or an increase of $V_{act} (p)$, respectively.
	Therefore, the inflection point $p_{infl}$ is being reflected by a minimum of $V_{act}(p)$.
	
	Herbst \textit{et al.} suggested that the inflection might arise from a nonlinearity of the volume with a change of pressure \cite{Herbst1993}, while Casalini and Bair attributed it to the pressure dependences of the compressibility and of the apparent activation energy at constant volume \cite{Casalini2008}.
	Bair emphasized the importance of the incorporation of this feature for physically meaningful models \cite{Bair2015} and proposed a combination of the McEwen-Model with a pressure equivalent \cite{Paluch1999} of the VF-law (Equ. \ref{equ-vf}) to give a phenomenological description:
	\begin{equation}
		\eta(p)=\eta_0 \times \left(1 + \frac{\alpha_0}{q} p \right)^q \times \textnormal{exp} \left(\frac{C_{F} p}{p_{\infty} - p} \right)
		\label{equ-hybrid}
	\end{equation}
	Here, $\eta_{0}$ is the viscosity at $p \rightarrow 0$.
	The fitting parameters $\alpha_{0}$ and $q$ originate from the McEwen equation, while $C_{F}$ and the divergence pressure $p_{\infty}$ are taken from the VF-like model. 
	Fits according to Eq. \ref{equ-hybrid} are presented in Fig. \ref{fig-meas} and, with identical parameters, in the inset.
	From the minima in the latter curves, the inflection pressures $p_{infl} (T=328\textnormal{\,K})\approx 350$\,MPa and $p_{infl} (T=350\textnormal{\,K})\approx 450$\,MPa can be derived.

	\begin{figure}[tb]
		\centering
		\includegraphics[width=8.5cm]{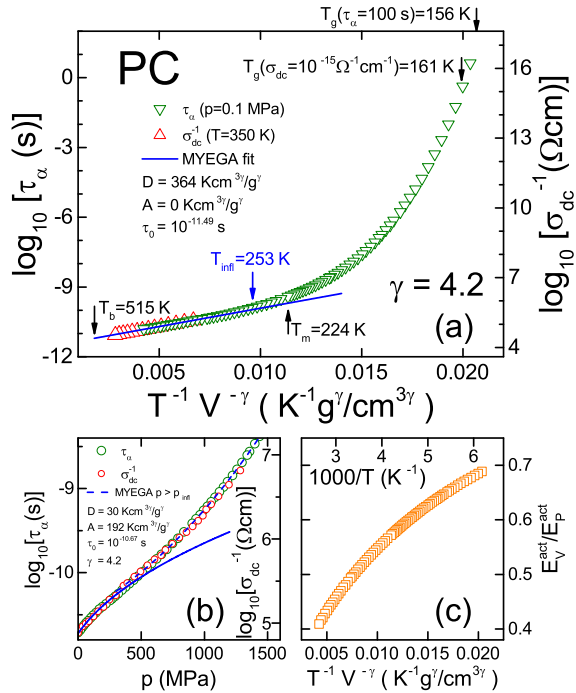}
		\caption{Ambient pressure relaxation times from Ref. \cite{Pawlus2004} versus scaling variable $\Gamma = T^{-1} V^{-\gamma}$ (a). In the low-$\Gamma$ range, conductivity data are added to expand the diagram nearly to the boiling point. The boiling, melting and glass temperatures $T_b$, $T_m$ and $T_g$ as well as the inflection temperature $T_{infl}$ (all at ambient pressure) are marked by the arrows. Fits of the low- and high $\Gamma$-regime data according to Eq. \ref{equ-myegatv} are given by the solid ($A = 0 \textnormal{ K}(\textnormal{g/cm}^{3})^{\gamma}$) and dashed ($A = 192 \textnormal{\,K}(\textnormal{g/cm}^{3})^{\gamma}$) lines, respectively. In (b), the isothermal pressure dependence of $\tau_{\alpha}$, generated from the density scaling criterion based on the data shown in (a) and pressure-volume-temperature measurements, is shown together with the measured $\sigma_{dc}$ 350\,K data from Fig. \ref{fig-meas}. The dashed and solid lines correspond to the fits in (a). The ratio of isochoric and isobaric activation energy is pictured in (c) as a function of $\Gamma$. The top scale gives the inverse temperature associated with $\Gamma$ at ambient pressure.}
		\label{fig-ds}
	\end{figure}

	The phenomenology of Eq. \ref{equ-hybrid} is, however, insufficient to gain a deeper insight into consequences of the dynamic crossover indicated by the inflection point. 
	To make progress towards a better understanding this phenomenon, we employ the density scaling idea in this study, as presented in Fig. \ref{fig-ds}(a).
	Here, the density scaling of the structural relaxation time $\tau_{\alpha} (\Gamma)$ is shown for the ambient pressure data taken from Ref. \cite{Pawlus2004}, with the addition of $\sigma_{dc}$ data from Fig. \ref{fig-meas} to expand the curve towards higher temperatures.
	It should be noted that the scales on the $\tau_{\alpha}$ and $\sigma_{dc}$ axis comprise the same amount of decades, but are shifted against each other to make the data comparable.
	This way, the thermodynamic evolution of dynamic properties can be presented from close to the glass transition to nearly the boiling point ($T_{g}=156$\,K and $T_{b}=515$\,K at 0.1\,MPa).
	To acquire the volume data, results of pressure-volume-temperature measurements and a Tait-like equation of state (EOS) reported earlier by Pawlus \textit{et al.} \cite{Pawlus2004} have been used.
	In the past, slightly different values of the scaling exponent have been reported \cite{Pawlus2004,Reiser2005,Casalini2008}.
	Indeed, we found $\gamma = 4.2$, in accordance with Reiser \textit{et al.} \cite{Reiser2005}, to get the best density scaling of the combined structural relaxation and dc-conductivity data.
	It can be seen clearly that the $\Gamma$ dependence changes from a slow increase at low $\Gamma$ to a much faster ascent closer to the glass transition.
	The crossover can be found approximately at $0.01 \textnormal{\,K}^{-1} \textnormal{g}^{\gamma}\textnormal{cm}^{-3\gamma}$ and $\tau_{\alpha} \approx 10^{-10}$\,s.

	It is worth noting that the scaling criterion (Eq. \ref{equ-scaling}) allows to generate numerically isotherms, isochores and isobars out of measured data as shown in detail in the supplementary information to Ref. \cite{Jedrzejowska2017}.
	Using this procedure with $\gamma = 4.2$ and the Tait EOS from Ref. \cite{Pawlus2004}, $\tau_{\alpha} (p)$-values were calculated based on the ambient pressure isobar shown in Fig. \ref{fig-ds}(a).
	The isotherm for $T = 350$\,K is presented in Fig. \ref{fig-ds}(b) exemplary, together with the according conductivity data from Fig. \ref{fig-meas} for comparison.
	The scales on both axis cover the same amount of decades, as above.
	Both curves show distinct accordance in the $p$-dependence, with negative curvature of the generated curve in the range from $\tau_{\alpha} = 2 \times 10^{-11}$\,s at ambient pressure up to the inflection point at $p_{infl} = 450$\,MPa and $\tau_{\alpha} = 8 \times 10^{-11}$\,s. 
	At higher pressures, up to a value of $\tau_{\alpha} = 3 \times 10^{-9}$\,s at 1400\,MPa, the increase in $\tau_{\alpha}(p)$ becomes faster than exponential.
	This correlation strongly supports the validity of data generation using the density scaling criterion.
	
	Of particular interest is the emergence of an inflection, concurring with the measured conductivity results, in the generated curve, while no such behavior was observed in the ambient pressure source data.
	The accordance of $\tau_{\alpha}(p)$ and $\sigma_{dc}^{-1}(p)$ becomes even more obvious in a comparison of the respective activation volumes.
	As can be seen in the inset of Fig. \ref{fig-meas}, both curves are in near perfect agreement over the whole pressure range, implying that the structural relaxation dynamics of PC can be followed by studies of the dc-conductivity.
	This is a promising result, since dielectric measurements of fast dynamics under high pressure are difficult experimentally, as opposed to conductivity measurements.
	
	Using the Tait EOS from Ref. \cite{Pawlus2004}, Eq. \ref{equ-myegatv} was adapted for the fitting of isotherms. 
	Separate fits of the $\tau_{\alpha}$ isotherm at $T=350$\,K for $p < p_{infl}$ and $p > p_{infl}$ can be found in Fig. \ref{fig-ds}(b).
	Below the inflection point, it yields $\tau_{0} = 10^{-11.49}$\,s, $D=364 \textnormal{\,K}(\textnormal{g/cm}^{3})^{\gamma}$ and $A = 0 \textnormal{\,K}(\textnormal{g/cm}^{3})^{\gamma}$.
	Using these results, $\tau_{\alpha}(\Gamma)$ can be described equally well for $\Gamma < 0.01 \textnormal{\,K}^{-1} \textnormal{g}^{\gamma}\textnormal{cm}^{-3\gamma}$.
	This is illustrated by the solid line in Fig. \ref{fig-ds}(a).
	It should be noted that a linear $\Gamma$ dependence, as it is given by $\tau_{\alpha} = \tau_{0} \times \textnormal{exp} \left( \frac{D}{T V^{\gamma}}\textnormal{exp} \left( \frac{A=0}{T V^{\gamma}} \right) \right)$, precludes the application of an Arrhenius law (Eq. \ref{equ-arrh}) to describe the data at high temperatures. 
	This coincides with the observations made on the $\sigma_{dc}$ isotherm at low pressures, showing that Eq. \ref{equ-parrh} is likewise improper. 
	As given by the dashed line in Fig. \ref{fig-ds} (b), Eq. \ref{equ-myegatv} can also be used to fit the results above the crossover point (\textit{i.e.}, $p > 450$\,MPa) with $\tau_{0} = 10^{-10.67}$\,s, $D=30 \textnormal{\,K}(\textnormal{g/cm}^{3})^{\gamma}$ and $A = 192 \textnormal{\,K}(\textnormal{g/cm}^{3})^{\gamma}$, implying that a single model is eligible to describe all measured data.
	
	To quantify the relative importance of the influences of temperature and free volume on a dynamic property, the ratio of the isochoric activation energy $E_{V,act} = R [\partial \textnormal{ln}(\tau_{\alpha}) / \partial T^{-1}]|_{V}$ and the isobaric equivalent $E_{p,act}$ can be calculated \cite{casalini2003}. 
	The ratio can take values between 0 and 1, indicating solely free volume activated and ideal thermally activated behavior, respectively.
	From Fig. \ref{fig-ds}(c) it can be seen that $E_{V,act}/E_{p,act}$ increases with higher values of $1/TV^{\gamma}$, which implies an increasing relative influence of the temperature over the volume on $\tau_{\alpha}$ \cite{Casalini2005}.
	Close to the boiling point, the volume is the more important parameter ($E_{V,act}/E_{p,act}=0.41$), while near the glass transition, a ratio of 0.69 is observed, \textit{i.e.} the thermal influence on the dynamics is roughly twice as high as the volumetric one.
	The increase of volumetric influence when approaching the boiling point is contradictory to purely energy activated behavior, supporting the employment of a non-Arrhenius model. 	
	
	\begin{figure}[tb]
		\centering
		\includegraphics[width=8.5cm]{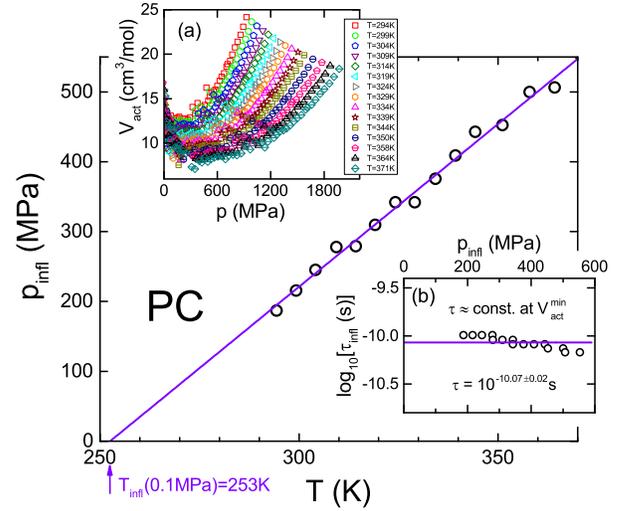}
		\caption{Temperature dependence of $p_{infl}$. The line is a linear fit of the data, extrapolated to the inflection temperature at ambient pressure (arrow). For reference, $V_{act}(p)$ is shown in inset (a) for $\tau_{\alpha}$ of selected isotherms. Inset (b) presents the relaxations times as a function of $p_{infl}$, where the line illustrates $\textnormal{log}(\tau_{\alpha}) = 10^{-10.07}$\,s.} 
		\label{fig-inft}
	\end{figure}
	
	By using a set of $\tau_{\alpha}$ isotherms generated as described above, the influence of temperature on the inflection point can be determined (see Fig. \ref{fig-inft}).
	$p_{infl}(T)$ is taken at the minimum of the corresponding $V_{act}(p)$ curves, shown in inset (a).
	As elaborated in Ref. \cite{Jedrzejowska2017}, measured base points are needed for the curve generation.
	Consequently, isotherms between $T=294$\,K and 371\,K were taken into account based on the underlying data collected at ambient pressure.
	In the given range, a linear increase of $p_{infl}$ from 187\,MPa at $T=294$\,K to 553\,MPa at 371\,K can clearly be observed.
	A linear $p_{infl}(T)$ dependence seems to be common behavior for materials without significant hydrogen bonding \cite{Thoms2017}. 
	A linear fit (line in Fig. \ref{fig-inft}) with a slope of 4.7 MPa/K is adequately describing the data.
	Extrapolation of this fit to ambient pressure results in an estimate of $T_{infl} (p=0.1 \textnormal{MPa}) = 253$\,K. 
	In Fig. \ref{fig-ds}(a), this temperature is marked by an arrow, and accords very well with the crossover from linear ($A=0\textnormal{\,K}(\textnormal{g/cm}^{3})^{\gamma}$) to faster ($A>0\textnormal{\,K}(\textnormal{g/cm}^{3})^{\gamma}$) growth of $\textnormal{log}_{10}\tau_{\alpha} (\Gamma)$.	
	Remarkably, $T_{infl}$ is nearly 30\,K higher than the melting temperature $T_m$, implicating a change of the molecular dynamics even before supercooling takes place.
	
	Another interesting result can be drawn when plotting the structural relaxation times at the inflection point $\tau_{\alpha,infl}$ against $p_{infl}$ (Fig. \ref{fig-inft} (b)).
	In fact, for all isotherms, $\textnormal{log}_{10}\tau_{\alpha,infl} \approx -10.07$ is almost constant, \textit{i.e.}, the inflection can be observed at isochronal conditions.
	
	Summing up, we performed isothermal high pressure dielectric measurements on propylene carbonate for two isotherms.
	Additionally, structural relaxation times for $T=350$\,K in the same pressure range were generated out of existing ambient pressure data employing the density scaling criterion and a Tait-like EOS.
	A direct comparison of the $\sigma_{dc}$ and $\tau_{\alpha}$ data interestingly shows a remarkable accordance of the pressure dependent behavior, suggesting that it is possible to access fast dynamics via conductivity measurements.
	Since direct measurements of $\tau_{\alpha}$ are difficult in high pressure set-ups in the fast dynamic range, this can offer a new route to expand spectra.
	
	A noteworthy observation is the presence of an inflection point, \textit{i.e.} a transition from slower to faster than exponential pressure dependence in both properties. 
	Similar observations were made before, both in PC and in other materials \cite{Cook1994,Casalini2008,Gacino2015,Thoms2017}.
	However, the presence of the inflection in the generated $\tau_{\alpha}$ isotherm is of special significance since no such feature can be observed in the reference isobar.
	This implies that it is possible to effectively predict high pressure properties via low pressure measurements when using the density scaling criterion.
	
	Our analysis of the high-$T$ resp. low-$\Gamma$ data indicates a clear deviation from an Arrhenius model regularly used \cite{Roland2005} to describe the fast dynamics. 
	This can be seen both in the linear $\Gamma$-dependence in Fig. \ref{fig-ds}(a) and the slower than exponential $p$-dependence in Fig. \ref{fig-meas} and \ref{fig-ds}(b) in the low $\Gamma$- respectively low $p$-ranges, as well as the increase of $E_{V,act}/E_{p,act}$ with $T$.
	Instead, we employed a modified MYEGA model (Equ. \ref{equ-myegatv}) to successfully fit the data, using the parameter $A=0\textnormal{\,K}(\textnormal{g/cm}^{3})^{\gamma}$, which reasonably indicates that the intact and broken constraints are energetically degenerated or the topological constraints are ineffective at $\tau < \tau_{infl}$ \cite{Mauro2009}.
	The same model with a positive value of $A$ can be used to appropriately describe the data at high $p$ range.
	Analyzing the temperature dependence of the inflection pressure $p_{infl}$, a linear increase with $T$ was observed.
	Besides, $\tau_{infl} \approx 10^{-10}$\,s is nearly constant for all curves, \textit{i.e.} the inflection point is found at isochronal conditions.
	We speculate that this phenomenon can be attributed to the transition from simple to cooperative molecular dynamics in glass formers.
	The crossover takes place at temperatures higher than the melting point, suggesting that the observed change of dynamics is a precursor of supercooling.

\begin{acknowledgments}
	S.P., M.P. and E.T. acknowledge the financial support of the project No. UMO-2015/17/B/ST3/01221 by the National Science Centre, Poland. 
\end{acknowledgments}

\bibliography{inflection}

\end{document}